# Predicting Social Media Engagement using Machine Learning

Ritwik Singh[1], Mayukh Majumdar[2], and Subodha Kumar[3]

[1] Interlake High School, Bellevue, Washington, USA
ritwikrule@gmail.com
[2] Knauss School of Business, University of San Diego, 5998 Alcalá Park, San Diego, CA 92110, USA
mmajumdar@sandiego.edu
[3] Fox School of Business, Temple University, 1810 Liacouras Walk, Philadelphia, PA 19122, USA
subodha@temple.edu

**Abstract.** Social media platforms are popular channels for disseminating information, owing to their large user bases and ease of access. Companies also use social media as an important aspect of the advertising process. By creating high-quality posts, companies can strengthen their engagement metrics and increase their follower count. While a growing body of research has examined social media engagement, fewer studies have jointly examined the visual, textual, and temporal features of image posts, even though these features collectively determine the performance of content on social media. To understand the important drivers of social media engagement, we collect image posts of furniture firms on Facebook and extract visual, temporal, and textual features from them using text and image analytics methods. We evaluate several machine learning models—including Random Forest, Light Gradient Boosting Machine (LightGBM), and eXtreme Gradient Boosting (XGBoost)—to assess the drivers and the prediction power of social media engagement using the features from our data. Our primary model, LightGBM, explains 81.8% of the variance in the engagement score, 2.3 times that of our baseline model. We observed that brand-relative deviation features, which capture how far a post departs from its brand's norms, are the strongest predictors, accounting for 51.3% of total importance. Our research quantifies the extent to which these features are associated with interactions and provides recommendations that organizations may consider.

## 1 Introduction

Social media platforms have become an important channel for advertising, with roughly 4.95 billion users worldwide and platforms such as Facebook reaching over 3.07 billion monthly active users [1-2]. Although much research and many practitioner guidelines focus on creating engaging content, user engagement is influenced by a complex set of textual, visual, temporal, and contextual factors, and social media managers continue to face challenges in creating content that performs well [3-4]. On platforms such as Facebook, users react not only to the message but also to characteristics of the post, including its visual content, design, and timing [5]. Recent advances in computer vision and machine learning offer new opportunities to study these dynamics, and accordingly, the aim of our study is to use machine learning to quantify the relationship between various feature groups and a post's engagement, and to rank the most important factors for brands to consider [5].

Throughout the course of our study, we use a weighted engagement score consisting of reactions, comments, and shares, the three most easily accessible engagement metrics, as our target variable. A weighted score is appropriate because a post's engagement is composed of multiple metrics, each of which requires a different amount of user effort and generates a different amount of visibility [6]. For social media managers, a weighted score provides a more actionable measure of a post's value, as it prioritizes high-effort metrics such as comments and shares, which generate more reach, while still accounting for reactions. We will explore how different feature groups impact post engagement, namely historical posting patterns, textual features, visual features, and temporal context. Studying these feature groups together is important, as social posts are a mixture of text, imagery, and posting time, and their interactions are what shape user engagement. Using a dataset of 32,456 Facebook posts from several different furniture companies, we utilize various machine learning algorithms to model the relationship between our feature groups and the target variable, analyze feature importance, and conclude with a set of recommendations for brands seeking to effectively utilize social media marketing.

Our dataset consists of firms from the furniture industry because furniture products are frequently showcased visually, and because firms increasingly rely on Facebook for marketing [7]. We employ the furniture industry as a representative empirical context rather than as the focus of our investigation. Furniture is a mature, saturated category in which established firms compete for infrequent, high-involvement, relatively expensive purchases largely through visual merchandising, which makes it a particularly appropriate setting for studying how visual and content-related features relate to engagement. A growing stream of research has leveraged machine learning, natural language processing, and, more recently, deep learning techniques to understand and predict social media engagement [8-9]. Prior studies have examined the role of textual, temporal, and content-based features in driving engagement, and seminal contributions have demonstrated the value of incorporating textual characteristics, sentiment, posting behavior, and visual content when predicting user reactions and post popularity [5, 10].

To address these gaps, our study makes two contributions. First, we combine visual, textual, and temporal features, extracted using deep-learning-based text and image analytics, within a single machine learning framework applied to an extensive dataset of furniture brands, whereas prior work has typically examined these feature types in isolation. Second, we introduce a set of brand-relative deviation features that measure how far a post departs from a brand's own historical norms, which prove to be among the most informative predictors of engagement. As a single-industry study, our contribution is best understood as an in-depth case study of a mature retail category rather than a general account of engagement drivers, and it offers practically relevant guidance for social media managers seeking to improve engagement.

## 2 Data, Variables, and Methods

We collected our data from Facebook, a leading online platform for advertising. We obtained a list of companies from furniture.com, a directory of furniture brands located in the United States. We identified 162 brands with active Facebook pages and collected a substantial number of posts from each brand. The final dataset consisted of 32,456 posts and included 249 features, including post content, comments, posting time, follower counts, and engagement metrics.[1]

After obtaining the necessary raw data, our goal was to develop a target variable that could be used for machine learning. As discussed earlier, a weighted score reflects that different engagement types carry different values for users. Accordingly, we construct a composite variable composed of the number of reactions, comments, and shares using a weighted sum, with reactions receiving a weight of 1, comments receiving a weight of 5, and shares receiving a weight of 10. We assigned comments and shares higher weights than reactions to reflect the effort required to create them and their impact on Facebook's recommendation algorithm. Aside from the real-world need to assign weights to different engagement metrics, our specific weights were motivated by established precedent for creating engagement scores [11]. After summing them together, we used a logarithmic transformation to account for variance.

Next, we discuss how we collect important aspects of social media posts that can explain engagement. We employ textual and image analytics methods and obtain textual, image, temporal, and post-related historical features.

**Textual Features**

- Linguistic and structural characteristics extracted using spaCy and standard text analytics, including word count, sentence count, lexical diversity, hashtags, emojis, capitalization, and binary indicators for questions, promotions, prices, and numerical content. These characteristics are motivated by prior work showing that longer, more readable Facebook posts with more hashtags tend to achieve higher engagement [12].
- Sentiment and emotion features obtained from TextBlob, Valence Aware Dictionary and sEntiment Reasoner (VADER), and Robustly Optimized Bidirectional Encoder Representations from Transformers Approach (roBERTa), capturing polarity, subjectivity, sentiment, and emotion-specific scores such as joy and sadness. These features are motivated by prior work

---

[1] Due to API rate-limiting constraints, we retrieved 200 posts per brand on average. To assess the effect of these incomplete captures, we collected additional posts for four representative brands (Ethan Allen Design, IKEA USA, Bassett Furniture, and Arhaus) and compared the limited and full samples. Key features differed by negligible amounts (for example, a 0.0043 difference in the logarithmic engagement score), and models trained on the limited and full samples produced similar results (an $R^2$ difference of 0.0114), indicating that our sample size was large enough that further additions did not meaningfully change the outcome.

showing that emotional content is a well-established driver of engagement and sharing [8, 13].

- Caption specificity, a score capturing how product-specific a caption is, measured by the density of industry-specific and descriptive terms, such as materials, product categories, and attributes, relative to caption length.

**Image Features**

- Low-level visual attributes extracted using OpenCV, including brightness, contrast, saturation, hue, aspect ratio, and color dominance. These attributes are motivated by evidence that higher-quality and more colorful images are associated with greater social media engagement [14].
- High-level semantic features generated using YOLO object detection (e.g., number of people, furniture, objects) and Contrastive Language-Image Pre-training (CLIP) embeddings. The presence and type of image content have likewise been linked to higher engagement [14].
- Additional CLIP embeddings with dimensionality reduction performed using Principal Component Analysis.

**Historical Post Features**

- Relative deviation of each feature from a brand's historical average, along with brand-specific standardized (z-score) measures quantifying how unusual a post is relative to the brand's historical distribution. Together these capture how far a post departs from a brand's established norms and are motivated by prior work on the role of visual and stylistic consistency in branded social media content [15].

**Temporal and Account Features**

- Temporal and account characteristics, including post age and a brand's monthly posting frequency. These are motivated by prior work linking posting activity to engagement [10].

We conducted feature selection using permutation importance and Shapley Additive exPlanations (SHAP) values [16], resulting in a final set of 172 informative features. This final set achieved performance similar to the full dataset, confirming that our feature elimination primarily removed noise. We constructed all features, including the brand-relative deviation and z-score measures, using training-set observations only, and then applied the fitted statistics to the held-out test set. We likewise performed feature selection within the training fold, so that no test observations informed either feature construction or selection.

We evaluated several machine learning models to identify the approach with the strongest predictive performance with respect to the weighted engagement score. Among the models considered, the LightGBM regressor achieved the best results and was therefore selected as our primary model. LightGBM is a gradient-boosting framework that builds an ensemble of decision trees sequentially and is known for its efficiency on high-dimensional data [17].

To benchmark performance, we also evaluated XGBoost and Random Forest, two additional tree-based ensemble methods, along with Ordinary Least Squares (OLS) regression as a baseline.

We divided the dataset using an 80–20 train-test split, resulting in 25,964 training and 6,492 test observations, and optimized the hyperparameters for all models using RandomizedSearchCV. For LightGBM, we used a column sampling rate by tree of 0.5, a learning rate of 0.08, a maximum depth of 7, a minimum of 50 child samples and 30 data points per leaf, 1,200 estimators, 47 leaves per tree, regularization alpha and lambda of 0.5, and a subsampling rate of 0.9. For XGBoost, we used a column sampling rate by tree of 0.7, a gamma of 0.2, a learning rate of 0.03, a maximum depth of 7, a minimum child weight of 10, 1,200 estimators, regularization alpha of 1 and lambda of 2, and a subsampling rate of 0.9. For Random Forest, we disabled bootstrap sampling and set no maximum depth, a maximum feature sampling rate of 0.4, a minimum of 2 samples per leaf and 10 to split an internal node, and 200 trees. Five-fold cross-validation (CV) was then conducted to assess the robustness and consistency of the results.

## 3 Results

Table 1 shows the model fit results, where we can observe that the LightGBM model had the highest $R^2$ after tuning and therefore performed the best. Although our baseline model (OLS regression) was only able to explain 35.44% of the variance, the tuned version of LightGBM improved substantially with an $R^2$ of 0.8178. Our model's predictive power was highly consistent, as its CV $R^2$ differed from the test $R^2$ by only 0.001, and the standard deviation across folds was 0.004.

**Table 1.** Overview of results using log-transformed engagement score

| Model | $R^2$ | RMSE | MAE |
|---|---|---|---|
| OLS Regression | 0.3544 | 1.4772 | 1.1674 |
| LightGBM | 0.8178 | 0.7847 | 0.5347 |
| XGBoost | 0.8121 | 0.7968 | 0.5641 |
| Random Forest | 0.8059 | 0.8099 | 0.5370 |

To ensure our findings were not a byproduct of natural growth with age, we re-estimated the models using three age-adjusted outcomes: a linear measure, a log-transformed measure, and a flexible curved fit, each estimated using only the training data. The model's predictive power remained relatively unchanged, with the $R^2$ staying between 0.81 and 0.82 and the top 15 features remaining mostly constant.

To test whether the consistency effect simply reflects larger, more established brands, we removed each brand's own average from both engagement and its deviation features, so that any surviving association must come from within-brand variation rather than brand size. The within-brand association persists after removing brand-level averages (Pearson's r changes from 0.029 to 0.047), suggesting that the relationship is not driven solely by persistent differences across brands.

To further quantify how much of our model's performance reflects brand identity versus content, we fit a model using only brand identity (the identity of the specific brand that authored a post) with no information about the post's content as a predictor. This brand-only model achieved an $R^2$ of 0.688. Since our full model reaches an $R^2$ of 0.8178, content features contribute roughly 0.13 in additional $R^2$, which indicates that while brand identity is a strong predictor of engagement, content features also add a significant share of predictive power within brands.

We also tested the deviation features against seasonality in three ways. Within each calendar month, all twelve deviation features kept their sign with only a small reduction in strength (about 6% on average). Removing each month's average engagement from the target and re-fitting produced nearly identical performance ($R^2$ of 0.8176 versus 0.8178), with all fifteen top features unchanged. Finally, models trained and tested separately within each season performed consistently ($R^2$ between 0.77 and 0.78) with stable rankings. Together, these results indicate that the deviation effects are only slightly affected by seasonal timing rather than produced by it.

**Table 2.** Feature group importance

| Feature Group | Number of Features | Total Importance | % of Total |
|---|---|---|---|
| Historical Posting Patterns | 53 | 0.4435 | 51.3% |
| Visual Features | 63 | 0.2799 | 32.4% |
| Textual Features | 50 | 0.1159 | 13.4% |
| Temporal Context | 7 | 0.0254 | 2.9% |

After testing the models and comparing their ability to explain the variance in our calculated engagement score, we also computed permutation feature importance for our feature groups. As shown in Table 2, a brand's historical posting behavior is the dominant group in explaining engagement, indicating that a brand's own posting history is strongly associated with its engagement. For reference, deviation is the difference between a value and the mean, while the z-score measures the number of standard deviations between the two, which we use as a more normalized alternative to deviation.

Several features were positively associated with engagement, including the raw hashtag count, which is the total number of hashtags in a post, and the image-count z-score, which is the number of images standardized relative to a brand's history. The z-score of the number of fully capitalized words was also positively associated with engagement, indicating that using more all-caps words than a brand's norm, often to emphasize announcements or promotions, is associated with higher engagement. The direction of each feature is determined from the sign of its mean SHAP value, which captures whether higher values of the feature push the model's predicted engagement up or down. This is distinct from permutation importance, which measures only how much a feature contributes to accuracy and not in which direction, so we rely on signed SHAP values to assign direction consistently across the figure and discussion.

The importance of these standardized deviation features suggests that posts that differ from a brand's usual style, such as those with an unusual number of hashtags or capital letters, tend to have less predictable engagement, further indicating that consistency with a brand's established pattern is associated with more stable engagement.

Additionally, although the first CLIP image component (CLIP image PC0) ranked as the third most important feature overall, we do not show it in Fig. 1 because it is a learned embedding that lacks a clear, interpretable meaning.

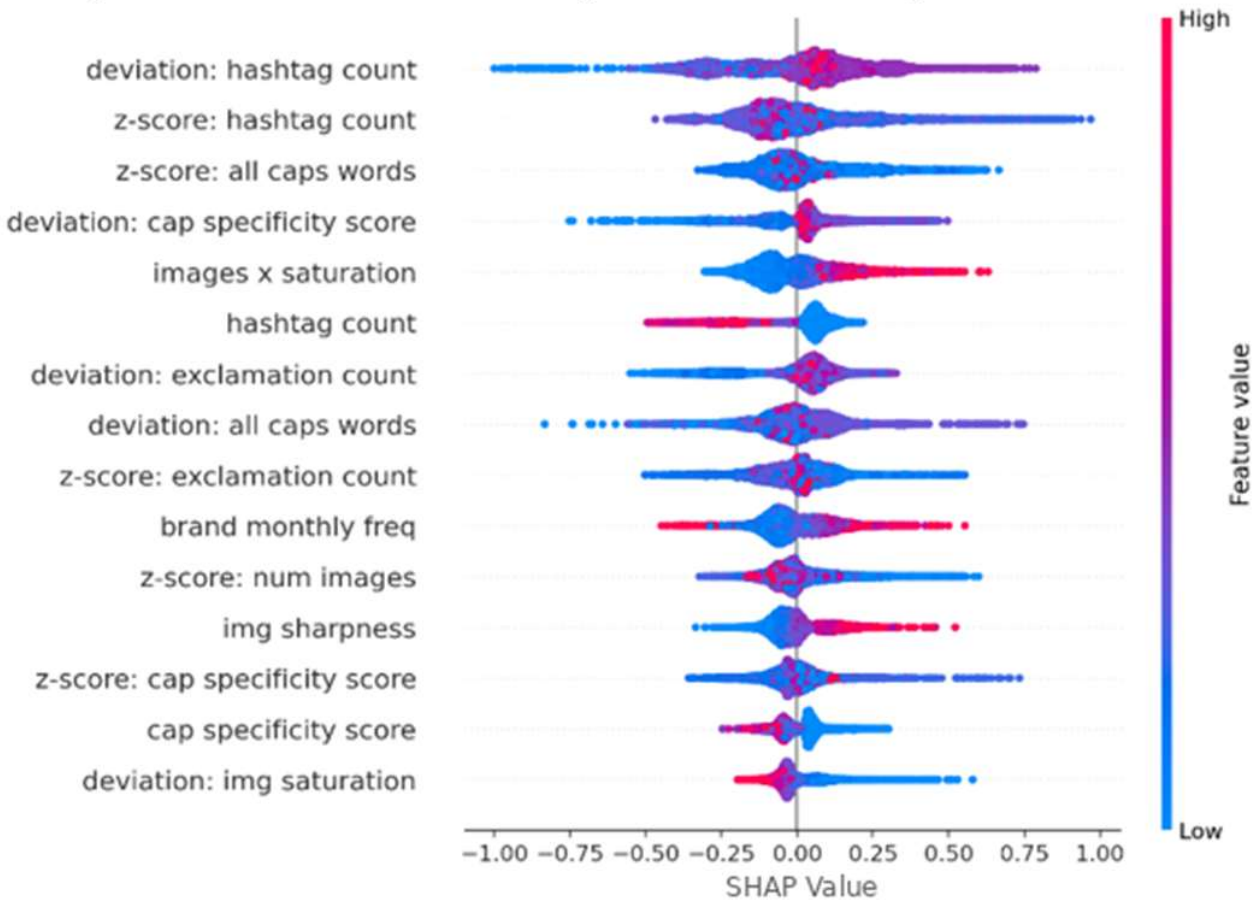

**Fig. 1.** Top SHAP features

As shown in Fig. 1, we also computed SHAP values for our features. We visualize these values using a beeswarm plot, in which each point represents a single post positioned by how much a feature pushed its predicted engagement up or down, and colored by whether the feature's value was high (red) or low (blue). For example, for the hashtag-count z-score, posts with an unusually high hashtag count (red points) tend to fall on the negative side, while posts with an unusually low hashtag count (blue points) tend to fall on the positive side. This indicates that posting far more hashtags than a brand typically uses is associated with lower engagement, whereas posting fewer is associated with higher engagement.

## 4 Discussion of Findings

To summarize, LightGBM performed the best when tuned, with an $R^2$ of 0.8178. We also obtained several feature importance rankings, which indicate which post characteristics are the strongest predictors of engagement levels. For example, Fig. 1 shows that hashtag-related features, capital letters, and exclamation marks are among the most important predictors of engagement, though the direction of these effects varies by feature. Our inclusion of state-of-the-art models such as XGBoost and LightGBM produced a substantial increase in the $R^2$ when compared to the baseline, along with a 46.9% decrease in the RMSE values. A paired t-test on the per-fold CV scores confirmed that LightGBM's advantage over XGBoost was statistically significant (t-statistic of 4.61, p-value of 0.010). Finally, our SHAP importance rankings reveal where our model's predictive power stems from, while the small difference between the CV $R^2$ and test $R^2$ demonstrates the model's stability and reproducibility across folds.

Based on the results of our study, there are several steps that social media managers can consider to improve the level of engagement that their posts receive, and by extension, their overall revenue and brand recognition. However, we note that the following recommendations are based on predictive associations rather than established causal effects, and that our model identifies features that accompany higher engagement rather than features guaranteed to increase it. Thus, managers may wish to treat these as suggestions worth testing rather than as a guaranteed method of raising engagement.

For example, within the historical feature group, one of the most prominent features was the hashtag count and its brand-relative measures. The raw hashtag count was positively associated with engagement. This is consistent with a study by Gkikas et al., which found that posts with more hashtags tended to perform better [12], possibly because hashtags help categorize posts and make them visible to interested users. The brand-relative z-score of the hashtag count, however, was negatively associated with engagement, a distinction that prior work, which has largely treated hashtag counts as absolute features [12, 18], has not fully captured. Together, these findings suggest that while using hashtags is generally beneficial, deviating sharply from a brand's established posting style is not. In contrast to this, the z-score for the number of images was positively associated with engagement, indicating that posting more images than a brand's norm is associated with higher engagement.

Some visual features also ranked highly. In particular, the interaction between the number of images and average image saturation was among the most important features, indicating that posts combining multiple images with vivid color tend to perform well. Image sharpness also ranked highly by importance, though its direction was mixed across posts rather than consistently positive. These findings suggest that visual presentation contributes to engagement, and managers may benefit from attention to lighting, color, and image selection when creating posts.

Textual features can also contribute to a post's engagement. Although longer captions are truncated, placing interesting information or eye-catching text at the beginning is a reasonable way for social media managers to encourage viewers to linger on their posts. The most important textual features were the number of words in solely uppercase letters, hashtag count, and our computed "specificity score" (calculated by counting how many industry-related words a caption uses). Readers tend to perceive words written in capital letters as a signal of importance, as they are often paired with important announcements (such as sales or new furniture announcements) to draw attention to posts. Similarly, our specificity score ranked highly in importance, although its brand-relative measures were negatively associated with engagement, suggesting that departing from a brand's usual level of caption specificity is associated with lower engagement.

Finally, we observed that a brand's monthly posting frequency was also a predictor of engagement. A brand's posting frequency is straightforward to adjust, and its association with higher engagement suggests that brands seeking to effectively promote their products online may want to maintain a consistent posting schedule. Aside from developing one's follower base, frequent community engagement and brand updates are associated with an enhanced brand image and may help attract more customers.

We also highlight some limitations of our study, along with suggestions for future research. First, brands are able to boost posts by paying to have them appear on Facebook users' homepages, and because our dataset did not contain this information, which is difficult to obtain, we cannot fully separate organic engagement from paid promotion. As a result, some outliers, particularly from smaller brands, might be explained by boosting in ways our model cannot confirm. Second, while we model the emotional content of captions, we do not explicitly capture the emotional content of the images themselves, which is a documented driver of engagement that methods such as context-based emotion recognition [13] could incorporate in future work. Third, because our study setting is the furniture industry, the generalizability of our findings to sectors with different products, audiences, and social media practices may be limited. However, the proposed framework is likely to be particularly relevant to other visually driven industries, in which a product's appearance and image quality are significant factors in purchase decisions.

Several directions could extend this work. Because video posts share many features with image posts, capturing a wider range of media types and comparing their engagement patterns would help optimize posting strategy. Future research should also examine other industries, especially those differing from furniture in promotional content, audience size, and demographics, to see how the patterns differ. Finally, although CLIP proved highly impactful, we did not decode it, and future work could measure how its components correlate with lower-level visual aspects to build actionable recommendations from CLIP embeddings and their principal components.

## 5 Conclusion

To summarize, we utilize machine learning on social media data to determine which factors are the strongest predictors of engagement and to create insights for brands seeking to increase their advertising's effectiveness within the furniture industry. After scraping over 32,000 Facebook posts from 162 furniture brands and using various Python libraries (such as YOLO and VADER) to obtain textual, temporal, visual, and historical features, we trained several different machine learning models. We compared a tuned version of LightGBM to two other advanced models, XGBoost and Random Forest, while using OLS regression as a baseline for performance comparisons. Ultimately, our strongest model (LightGBM) explained 81.78% of the variance. Additionally, we found that features such as the deviation and z-score of the hashtag count, image saturation, and the number of images were among the most important predictors. Features such as the number of exclamation marks and their brand-relative measures were also predictive. From these results, we suggest that social media managers seeking to improve their posts' engagement levels may want to consider using high-quality images and supporting them with eye-catching captions. Finally, posts using hashtags showed a strong positive association with engagement, consistent with their role in helping Facebook's algorithm recommend content to a wider audience. This suggests that increased usage of hashtags could be a strategy worth testing for managers seeking to broaden a post's reach.